# RNA-binding antiterminators:
# Regulation of metabolism and pathogenicity in bacteria

D. Soussan[a], A. Tahrioui[b], RR. de la Haba[c], A. Forge[b], S. Chevalier[b], O. Lesouhaitier[b]* and C. Muller[a]*

[a] *Univ Caen Normandie, Normandie Univ, CBSA UR 4312, F-14000 Caen, France;*

[b] *Univ Rouen Normandie, Normandie Univ, CBSA UR 4312, F-76000 Rouen, France;*

[c] *Department of Microbiology and Parasitology, Faculty of Pharmacy, University of Sevilla, 41012 Sevilla, Spain*

*corresponding authors. These authors contributed equally to this work.

Diane Soussan:
diane.soussan@gmail.com
0000-0003-4190-3567

Ali Tahrioui:
ali.tahrioui@univ-rouen.fr
0000-0003-4064-820X

Rafael R. de la Haba:
rrh@us.es
0000-0002-4615-780X

Adrien Forge:
adrien.forge@univ-rouen.fr
0009-0008-1641-9881

Sylvie Chevalier:
sylvie.chevalier@univ-rouen.fr
0000-0001-6489-438X

Olivier Lesouhaitier*:
olivier.lesouhaitier@univ-rouen.fr
0000-0001-8235-8919

Cécile Muller*:
cecile.muller@unicaen.fr
0000-0003-3078-1920

**Abstract**

Antiterminators are essential components of bacterial transcriptional regulation, allowing the control of gene expression in response to fluctuating environmental conditions. Among them, RNA-binding antiterminator proteins play a major role in preventing transcription termination by binding to specific RNA sequences. These RNA-binding antiterminators have been extensively studied for their role in regulating various metabolic pathways. However, their function in modulating the physiology of pathogens requires further investigation. This review focuses on RNA-binding proteins displaying CAT (Co-AntiTerminator) or ANTAR (AmiR and NasR Transcription Antitermination Regulators) domains reported in model bacteria. In particular, their structures, mechanism of action, and target genes will be described. The involvement of the antitermination mechanisms in bacterial pathogenicity is also discussed. This knowledge is crucial for understanding the regulatory mechanisms that control bacterial virulence, and opens up exciting prospects for future research, and potentially new alternative strategies to combat infectious diseases.

# Introduction

RNA polymerase (RNAP) is a multi-component enzyme that carries out transcription in all living organisms. After the initial recognition of a gene promoter region, sigma factor-assisted RNAP transcribes the first 10 to 15 nucleotides before the sigma factor is released to initiate elongation. RNAP then transcribes the DNA until it encounters termination signals, which cause rapid and irreversible dissociation of the nascent RNA transcript from RNAP. Transcription termination is caused by two main processes, one of which relies on the helicase Rho protein to remove RNAP (Rho-dependent termination), while the other one depends on the RNA sequence itself (intrinsic termination), and is referred to as Rho-independent termination[1]. In the latter, RNAP stops at a DNA sequence characterised by a GC-rich hairpin followed by a poly-T strand, leading to its dissociation from the DNA template without the help of auxiliary factors [2].

In bacteria, most transcription termination occurs *via* intrinsic RNA terminators located either at the end of genes to prevent transcription of the downstream genes [1], or in the upstream regulatory leader regions (or untranslated regions [UTR]) to adjust gene expression in response to metabolic and/or environmental signals [3]. Since RNA hairpin formation is mainly a passive process in bacteria, transcriptional antitermination is considered to be an important regulatory mechanism that allows RNAP to bypass or prevent the folding of hairpins and thus control gene expression [4].

Bacteria have several pathways in which RNAP can bypass a terminator, most of which involve the formation of alternative hairpins. One of these is a passive mechanism discovered in bacteria that is related to the action of stalled ribosomes and is known as termination attenuation. This mechanism gathers transcription and translation, and was described in *Escherichia coli* for the *trp* operon during tryptophan (Trp) starvation. When the concentration of Trp-tRNA is low, the ribosome stops at one or both Trp codons of the leader peptide while the RNAP continues transcription. These conditions favour the formation of an antiterminator secondary structure, allowing transcription of all the genes in the *trp* operon to continue. Otherwise, in the presence of a high concentration of Trp RNA, transcription of the leader peptide proceeds smoothly up to its stop codon. Progress of the ribosome, synchronised with that of the RNAP, prevents the formation of the antiterminator structure and promotes the formation of a transcription terminator. When this structure is formed, RNAP ceases to function prematurely, and the elongation complex dissociates before it reaches the first gene of the *trp* operon [5]. Another antitermination mechanism involves small molecules and mRNA leader riboswitches that undergo structural rearrangements in response to variations in ion concentration or upon binding to small compounds such as purine, flavin mononucleotide, or ppGpp for instance [6].

The formation of alternative RNA structures can also be modulated by RNA-binding proteins, leading to termination or elongation, and enabling the antitermination mechanism. In some bacterial RNA-binding antiterminators such as BglG and SacY found in *E. coli* and *Bacillus subtilis*, respectively, the CAT (Co-AntiTerminator) domain  seems to be conserved and used in regulation of sugar utilization operons (Figure 1A) [7,8]. The CAT domain binds to specific RNA motifs known as ribonucleotidic antiterminator (RAT) preventing the formation of terminator structures in nascent mRNA and allowing transcription to continue [9–11]. In many antiterminator proteins, the CAT domain which occur usually at the C-terminus is followed by regulatory domains, such as phosphotransferase system (PTS) regulatory domains (PRDs) (Figure 1A). These regulatory domains play a critical role in modulating the activity of the CAT domain. Most of the BglG/SacY antiterminators are involved in metabolism, and possibly also in virulence, particularly in opportunistic bacteria [12]. Another RNA-binding domain found in bacterial transcription antiterminator proteins is the ANTAR domain (AmiR and NasR

Transcription Antitermination Regulators) (Figure 1B), which bind RNA hairpin motifs within the nascent RNA chain to prevent termination [13,14]. This domain has been detected in several two-component system (TCS) response regulators and, unlike the majority of TCS output domains, binds RNA rather than DNA [15]. ANTAR domains are modular and include several types of sensors, or receptor domains, such as cGMP-related phosphodiesterases, adenylyl cyclases, FhlA (GAF) and Per-Arnt-Sim (PAS) domains (Figure 1B) [13,16]. As a result, proteins with ANTAR domains have been involved in the regulation of many bacterial metabolic and regulatory processes [16–18].

All these transcriptional antitermination mechanisms that occur in the 5'UTR have two advantages: (i) they avoid unnecessary energy consumption in the expression of regulated genes, and (ii) they enable the bacteria to respond quickly and efficiently to changes in environmental conditions, such as the availability of carbon sources.

This review summarises the current knowledge of characterised RNA-binding antiterminators displaying mainly RNA-binding CAT or ANTAR domains at the level of their structure, mode of action, regulation, and identified targets in model microorganisms (Table 1). It also highlights their functions in cellular processes, with particular emphasis on their involvement in bacterial metabolism, and pathogenicity in terms of infection, virulence, and biofilm formation.

## Antiterminators with RNA-binding CAT domain

Antiterminators with CAT domain are mainly found in *Bacillota*, followed by *Pseudomonodota* and *Actinomycetota*, as illustrated on Figure 1A. The activity of these antiterminators is always dependent of regulatory domains related to cell metabolism, and more specifically two PRD domains, in more than 95% of the analysed sequences. However, antiterminator with CAT domains are very rare in other bacterial phyla. The most extensively studied antiterminator in this family is the BglG protein in *E. coli*, the structure and the mechanism of which are described in the next section.

### *The BglG antiterminator in* Escherichia coli

In *E. coli*, the *bgl* operon consists of three genes: *bglG*, *bglF* and *bglB* (Figure 2), which encode functions essential for regulating the uptake and degradation of β-glucosides. These polysaccharides are composed of disaccharide units linked by covalent β-glycosidic bonds [19,20]. BglG is a transcriptional antiterminator that regulates the expression of its own operon, encoding BglG, the PTS enzyme EIIBCA BglF, responsible for β-glucoside transport, and the phospho-β-glucosidase BglB, which hydrolyzes the phosphorylated carbohydrates [7,19,20].

### *BglG structure*

The BglG antiterminator family is a group of regulatory proteins with an inducer-dependent phosphorylation activity, first identified in *E. coli* [9,21,22]. They are composed of an RNA-binding domain, corresponding to the so-called CAT domain, and two PRDs, called PRD1 and PRD2. BglG exists in monomeric and dimeric forms, the latter being initiated by PRD2 form the RNA-binding domain [23,24]. The *E. coli* BglG protein has not been crystallized, unlike its *B. subtilis* counterpart SacY, which is described in the next section.

### *BglG antitermination mechanism*

In the absence of β-glucoside, transcription of the *bgl* operon is initiated by the RNAP, which rapidly encounters a Rho-independent transcription terminator located in the untranslated region (5'UTR) of the *bglGFB* operon (Figure 3). BglG is indeed phosphorylated by BglF and remains inactive[7,19]. In the presence of the inducer sugar, BglF transports this carbon source and transfers its phosphate group to β-glucosides, resulting in the dephosphorylation and dimerization of BglG, and the formation of an active RNA-binding domain [21,25–27]. When active, BglG binds to a 30-nucleotide palindrome RAT RNA sequence, located in the *bgl* operon leader region. This sequence adopts a stem-loop structure that partially overlaps the intrinsic terminator [23].Thus, BglG bound to the RAT sequence then prevents terminator formation, and allows the RNAP to progress through the downstream coding gene sequence [7,9,22]. BglG activation depends on the phosphorylation state of the regulator, for which two alternative hypotheses have been proposed. While the first one suggests that BglF phosphorylates the His-208 on the PRD2 domain [28,29], the second suggests that this is the site of positive phosphorylation mediated by the histidine-phosphorylatable phosphocarrier protein (HPr), while BglF phosphorylates the His-101 on the PRD1 domain [30].

*Other levels of regulation*

HPr-mediated phosphorylation of BglG is thought to be a shortcut route to control the flux of other PTS sugars and their coordinated utilization [27]. The *E. coli bgl* operon is also subject to carbon catabolic repression (CCR) *via* the CRP protein (cAMP receptor protein, also known as CAP for catabolite activator protein) and the glucose-specific enzyme EIIA[Glc] (Figure 3). The CRP-cAMP complex, which is known to be involved in the expression of the PTS components, facilitates the expression of the *bgl* operon in absence of glucose and in presence of β-glucosides. EIIA[Glc] stimulates the phosphorylation of BglG by BglF in presence of glucose to cause a drastic decrease in the operon transcription. Glucose also interferes with BglG activity, by modulating BglF phosphorylation and thus its activity [26,31].

To keep BglG inactive in the presence of glucose or in the absence of β-glucosides, the antiterminator is phosphorylated by BglF and sequestered at the membrane. The presence of β-glucosides in the growth medium results in rapid dephosphorylation of BglG [28], allowing its release by BglF before it is attracted to the cell pole through interactions with Enzyme I (EI) and HPr [32,33]. A PRD1/PRD2 interaction also occurs when BglG is in its monomeric form, to limit dimer formation, and acts as a reservoir. This interaction is supported by BglF, whereas the presence of β-glucosides favours the dimeric form [24,34]. Interestingly, it has been shown that when BglG is artificially anchored to the inner membrane, it remains active without migrating to the transcription site [35].

Dole *et al*. have shown that *bgl* operon expression is also regulated by other players, such as the H-NS protein [36,37]. H-NS is an important pleiotropic regulator of bacterial physiological adaptation to external signals. This protein enhances the repression of the *bgl* operon expression when cellular transcription rates are low. The H-NS can bind to DNA sequence located upstream the *bgl* promoter and/or in the *bgl* operon to form a DNA hairpin. This stem-loop either prevents RNAP from binding to the promoter region or traps this enzyme into the hairpin, thereby preventing transcription of mature *bgl* mRNA [38]. By its binding on DNA downstream from the promoter, H-NS was also shown to block RNAP elongation and to induce transcription termination[37].

*Cellular processes involving BglG*

As a regulator, BglG controls the expression of the *bgl* operon, but it also has a major impact on bacterial gene expression. For example, Gordon *et al*. searched for RAT-like motifs in the genomes of K12 and uropathogenic *E. coli*, and found that most of the putative RATs do not overlap with terminators. Their results suggest that the BglG antiterminator is involved in mechanisms other than sugar utilisation and may have activities other than antitermination [39].

It has been shown that BglG remains bound to RNA even after performing its antiterminator role, and then enhances the stability of the 5′UTR of the *bgl* mRNA, by blocking accessibility to ribonucleases. In addition, in the presence of BglG, the 5′UTR of the *bgl* operon adopts an alternative secondary structure that protects it from endonucleases [40]. In *E. coli*, the regulator also binds to RNAP and more specifically to the β'-subunit to facilitate elongation and ensure complete transcription of the operon. To do this, it was suggested that BglG could interact with RNAP and move with it beyond the antitermination site, promoting transcription elongation [41].

A proteomic study in *E. coli* showed that BglG regulates at least 12 target genes [42]. Among them, BglG affects the expression of *gcvA* by destabilizing its mRNA and preventing its translation. This gene encodes a regulator of a non-coding RNA called *gcvB*, which plays a role in the expression of the oligopeptide transporter OppA. Through this intermediary, BglG is thought to have an indirect effect on stationary phase growth. BglG is also involved in the expression of *gadE*, which encodes a key regulator of pH homeostasis, and lipopolysaccharide (LPS) biosynthesis. GadE regulates the expression of the *ridA* and *lrp* genes, encoding a deaminase and a leucine-sensitive regulatory protein, respectively. The BglG protein binds RAT sequences upstream of *gcvA* and *gadE*, without the involvement of a terminator. Thus, the BglG regulator has a pleiotropic role, with a strong effect on cell physiology, independently of its antiterminator function and in the absence of the inducing sugar [43,44].

The BglG antiterminator has also been shown to play a key role in the pathogenesis of *E. coli*. In particular, the *bgl* operon has been shown to be induced *in vivo* during sepsis (Table 1) [45]. These results raise questions about the importance of these genes during infection and their presence in pathogens. For example, a complete *bglGFB* operon has been identified in *Shigella sonnei*, whereas bacteria such as *Salmonella* spp*., Proteus mirabilis*, *Enterobacter aerogenes,* and *Pseudomonas aeruginosa,* have only the *bglG* homolog [45]. BglG has also been shown to be involved in the virulence of *Listeria monocytogenes* [46]. Indeed, *bglG* mutation severely impairs bacterial invasion and reduces virulence in mice (Table 1). The authors suggest that *bglG* is involved in the sensing and consuming of carbohydrates from the host cell during infection, thereby facilitating the pathogenesis of *L. monocytogenes*. Thus, BglG is a well-studied regulator with a pleiotropic role that is not limited to metabolism or transport, but also in membrane homeostasis and virulence.

### The SacT/SacY/LicT antiterminators

Although the *bglG* system was first described in *E. coli*, such systems are highly conserved in bacteria and are also present in Gram-positive bacteria, such as SacT, SacY and LicT in *B. subtilis* [7,47,48]. The SacT/SacY/LicT antiterminators were extensively studied as part of the antitermination mechanism in *B. subtilis* during pioneering studies, which then served as a model for works on pathogens.

#### SacT/SacY/LicT structure

The SacT/SacY/LicT RNA-binding antiterminators have a common structure, and like BglG, they have two functional domains: the CAT domain involved in RNA binding, and the regulatory PRD domains.

*CAT domain.* The CAT domain of SacT/SacY/LicT proteins is located at the N-terminus and consists of 55 amino acids. Each monomer is composed of a four-stranded antiparallel β-sheet, with tight turns connecting strand 1 to strand 2 and strand 2 to strand 3, and a long hairpin connecting strand 3 and strand 4 (Figure 4A) [49]. The dimers are stabilised by the orthogonal stacking of the two β-sheets, forming a β-barrel closed on one side by four main-chain hydrogen bonds between the two β4-strands of the two monomers. On the other side, each barrel is closed by one symmetrical hydrogen bond. While one side of the β-strands forms the dimer interface, the other face is in contact with a hydrophobic cluster of residues from the long stem-loop between the third and the fourth β-strand, forming the hydrophobic core of the monomer [23,49].

*PRD domains.* The regulatory region consists of two structurally similar domains corresponding to PRD1 and PRD2 [50]. Each domain forms a compact bundle of five helix (α1 to α5). The PRD module contains a core consisting of two pairs of antiparallel helix. The first pair contains the antiparallel helix α1 and α4, while the second pair contains α2 and α5. The third helix (α3) is perpendicular to α5 at the periphery of the bundle (as LicT, Figure 4B). In its dimerised form, the phosphorylation sites of PRD2 are inaccessible to enzymatic partners. The PRDs undergo significant rearrangements between the inactive and the active forms, in particular the PRD2 which is highly mobile in the inactive state, and locked upon activation by phosphorylation [51]. Thus, in the inactive state, the PRD2 undergoes a large movement resulting in a dimer opening that exposes the phosphorylation sites. This movement provokes additional structural rearrangements of the PRD1 interface and the CAT-PRD1 linker.

*The SacY antiterminator*

In *B. subtilis*, SacY is a transcriptional antiterminator involved in the expression control of *sacB* and the *sacXY* operon (Figure 2) [47,52,53]. SacX is a sucrose-specific EIIBC<sup>Suc</sup> PTS transporter, and SacB is a levansucrase involved in sucrose degradation [47,52,54,55]. Identification of the regulatory site upstream of the *sacB* gene revealed that the regulatory pattern is similar to that of *bglG* [8,22,56–58]. Surprisingly, examination of the RAT sequence of the *sacXY* operon revealed the presence of an imperfect palindrome, corresponding to a long RNA hairpin structure, which has been suggested to be a Rho-independent terminator [53,59,60].

SacY activity has been shown to be independent of HPr [61], but, conversely to SacT, negatively regulated by His-99 phosphorylation [8,55]. Cross-regulation between SacY and SacT has been reported, as these two antiterminators are involved in sucrose utilization and exhibit very similar regulation. [10,62]. The expression of *sacXY* has also been shown to be affected by the DegS-DegU TCS (formerly *sacU*), a regulatory system with pleiotropic regulatory roles [47,53].

In *E. faecalis*, the NagY protein, which share 36% homology with SacY, has been shown to play a central role in metabolism and virulence (Table 1). The *nagY* gene forms an operon with *nagE*, which encodes an *N*-acetylglucosamine-specific EIICBA PTS transporter [63]. The RAT sequence found in the 5'UTR of the operon is an imperfect inverted repeat suggesting a regulation similar to that of SacY [12,22,64]. NagY is able to modulate the expression of another gene, *hylA*, which encodes a hyaluronidase involved in virulence in the *Galleria mellonella* model, biofilm formation, the degradation of glycosaminoglycans (GAGs) [12], and urinary tract infection [65]. GAGs are macromolecules found primarily in the extracellular matrix and mucus of the host. Their degradation has two main benefits: it reduces the viscosity of the host cell surface, and it provides a source of nutrients, such as *N*-acetylglucosamine, that facilitates pathogen growth during infection.

*The LicT antiterminators*

In *B. subtilis*, LicT is involved in the regulation of β-glucan utilisation, such as lichenan or salicin, and regulates the expression of the *licTS* (Figure 2) and *bglPH* operons [66]. The *licS*, *bglP* and *bglH* genes encode an extracellular β-glucanase, a β-glucoside-specific EII transporter, and a phospho-β-glucosidase, respectively [11,66,67]. LicT shares approximately 40% of sequence identity with its *E. coli* homolog BglG or other BglG/SacY family antiterminators [67,68].

LicT has been shown to be positively regulated by HPr [11,69]. Its activation is dephosphorylated at His-100 of PRD1, which is the target of BglP, and phosphorylated at His-207 and His-269 of PRD2 by HPr (Figure 4B) [49,64,70]. The allosteric propagation of the phosphorylation signal from the PRD domains to the PRD1-CAT linker ensures the coordinated transmission of structural information between the regulatory modules and the RNA-binding domain[71]. Although LicT's CAT and PRD1 domains appear stable in all active states, differences are observed in the linker region, suggesting that it plays a significant role in signal transduction[72]. Like BglG, the LicT phosphorylation by PTS components induces dynamic subcellular localisation at the subpolar level in the presence of an inducer such as salicin [73]. The expression of the *bglPH* operon has also been shown to be affected by CCR through the presence of a *cre* site [67]. Mutants deleted for *cre* and RAT/terminator sequences are completely independent of CCR, whereas *cre* mutations caused only partial relief [11]. This suggests a complementary regulation of LicT and CCR. A second CCR-linked mechanism is indirectly mediated by P-Ser-HPr, which is unable to phosphorylate (and thus activate) LicT [11,69,74]. Consistent with this concept, Lindner *et al*. have shown that a mutant insensitive to HPr phosphorylation is no longer susceptible to CCR[75].

To identify other potential LicT targets, Gordon *et al*. searched the *B. subtilis* genome for RAT-like motifs and identified 11 putative RAT sequences, in addition to those known for the antiterminator operons. Similar to *E. coli*, most of the identified putative RATs do not overlap with a terminator, suggesting that these elements do not play a role in antitermination[39].

In *S. pyogenes*, LicT regulates the *bglPB* operon, which is involved in salicin metabolism. *bglP* encodes a β-glucoside-specific PTS EII and *bglB* encodes a phospho-β-glucosidase. BglB hydrolyses β-glucosides into glucose and salicylic alcohol by specifically cleaving the β-linked glycosidic linkage between glucose residues. The *licT-bglPB* operon is subjected to CCR, as the operon is repressed in the presence of high glucose concentrations while induced with low glucose concentrations. Deletion of *bglP* and *bglB* result in growth defects of group A streptococci in the presence of sugars such as fructose and mannose, as well as deregulation of virulence-related genes involved in blood dissemination, biofilm formation, streptolysin S-mediated haemolysis and localized ulcer lesion formation [76]. Similarly, LicT-dependent β-glucoside metabolism plays a central role in *in vitro* adhesion, biofilm formation, growth, and *in vivo* colonization in *Streptococcus gordonii* (Table 1) [77].

## Antiterminators with RNA-binding ANTAR domain

Unlinke the CAT domain family of antiterminators, which is highly homogeneous among bacteria (Figure 1A), proteins with ANTAR domains can be highly diverse. Indeed, the ANTAR RNA-binding domain can be found alongside a variety of different regulatory domains (Figure 1B). Most of these regulators are associated with specific response regulator receiver domains for all the phyla. This receiver domain enables the response regulator protein to be activated in response to a specific stimulus. The other regulatory associated domains are also activated in response to environmental signals, as described in the following sections of this review. As illustrated in Figure 1B, most of the ANTAR domains are found in *Actinomycetota* and *Pseudomonodata*.

### *The AmiR antiterminator in* Pseudomonas aeruginosa

AmiR is an antiterminator expressed in all *P.s aeruginosa* strains and encoded in the *ami* operon. This operon was first described and characterised by Clarke's team [78], and its products enable *P. aeruginosa* to metabolize a number of aliphatic amides to use them as carbon and nitrogen sources *via* the tricarboxylic acid (Krebs) cycle [79].

The *ami* operon comprises 4 to 6 genes, depending on the *P. aeruginosa* strain (Figure 5A). In PAO1 strain, the operon consists of 5 genes, *i.e. amiS*, *amiR*, *amiC*, *amiB* and *amiE* (annotated PA3362 to PA3366, respectively) (*www.pseudomonas.com*) (Figure 5A and 5B). By contrast, strain PA14 only expresses *amiR*, *amiC*, *amiB*-like, and *amiE* (annotated PA14_20560 to PA14_20590, respectively) (*www.pseudomonas.com*) and lacks *amiS* (Figures 5A and B). Note that in both strains, the *ami* operon begins with a 'leader' sequence, sometimes referred to as the *amiL* open reading frame or the *amiE* leader (PA3366.1, Figure 5B). This non-coding RNA enables ribosome binding and contains a binding site for the Hfq chaperone [80]. The first characterised function of the *P. aeruginosa ami* operon was the conversion of short-chain aliphatic amides to their corresponding organic acids, an activity supported by the AmiE amidase, which hydrolyses aliphatic amides allowing *P. aeruginosa* to recover carbon and nitrogen sources. The *amiR* and *amiC* genes encode the positive and negative regulators of the *ami* operon, respectively [81]. The *amiB* and a*miS* genes are poorly described in the literature: AmiB has been hypothesised to function as a chaperone, and when coupled with AmiS, it may be involved in an ABC-type transport system [82].

*AmiR structure*

AmiR has been shown to form a dimer by binding to a protein consisting of 196 amino acids (Figure 5C, PDB 1QO0) [83]. The monomer consists of an N-terminal globular domain, a central coiled-coil region and a C-terminal helical domain. The presence of a RNA-binding ANTAR domain is predicted from the amino acid 135 to the amino acid 190. The AmiC–AmiR complex has been shown to crystallize with one AmiR dimer and two AmiC monomers in the asymmetric entity (Figure 5C) [83]. This crystal structure allowed the identification of the AmiC-dependent regulation mechanism of AmiR by sequestration rather than phosphorylation. When the AmiC subunits interact with AmiR, they interfere with the accessibility of the ANTAR domain, preventing it from interacting with the RNA (Figure 5C).

*AmiR antitermination mechanism*

The activation of the *ami* operon is induced by the presence of amides, such as acetamide, propionamide or lactamide (Figure 5A). Conversely, its expression is repressed by formamide and butyramide, which are considered antagonists of the AmiC sensor [79,84]. Recently, other natural compounds have been identified as AmiC agonists. Natriuretic peptides have been shown to bind purified AmiC. C-type natriuretic peptide (CNP) [85] and human atrial natriuretic peptide (hANP) [86] have been shown to bind purified AmiC with a quite similar affinity ($K_D$ of 2 µM for CNP; $K_D$ of 5 µM for hANP). In addition, CNP strongly activates transcription of the entire *ami* operon, suggesting that the CNP-AmiC complex releases the AmiR regulator, thereby allowing its antiterminator activity [85].

It has been shown that AmiR activates the expression of the *ami* operon. Overexpression of AmiR leads to increased amidase activity, and to constitutive expression of AmiE [87]. It was then discovered that the operon leader sequence contains a Rho-independent transcription terminator [88]. Deletion of this stem-loop also resulted in constitutive expression of AmiE suggesting that AmiR acts as a transcriptional antiterminator, a hypothesis that is supported by the sequence similarities between the binding sites of BglG and AmiR [89].

AmiC sequesters AmiR, thereby inhibiting its activity as a transcriptional antiterminator (Figure 5C). To release AmiR and to enable its activity, an agonist such as acetamide must bind to AmiC to remove its anchoring to AmiR [81,84,90] and allow expression of the *ami* operon [83,91].

Under inducing conditions that result in the release of a significant amount of free AmiR (a 27.3-fold increase in quantity), *amiE* is the most abundant transcript of the *ami* operon [90,91]. As being the first gene positioned after the first hairpin (so called T1), *amiE* is strongly transcribed once this terminator is bypassed, while *amiB* shows lower transcription levels, and *amiC* is very poorly transcribed. This control of *amiC* expression avoids limiting AmiR activity [91]. Therefore, the AmiR target is the mRNA leader sequence of the *ami* operon, and its binding to this sequence prevents specific hairpin structure, allowing RNAP activity [17]. Norman *et al*.

found that the stoichiometry of the interactants AmiR dimer and RNA was 152:1, corresponding to a $K_D$ value of 1.1 nM [90].

*Cellular processes involving AmiR*

Since these pioneering studies, little has been reported in the literature about the regulation of the *ami* operon until the discovery 10 years ago that a family of human hormones, the natriuretic peptides, are capable of directly binding to the AmiC sensor. This binding and the subsequent release of AmiR greatly altered the ability of *P. aeruginosa* to form biofilm [85,92], and triggered mature biofilm dissemination. This suggest that the antiterminator AmiR is involved in *P. aeruginosa* virulence, in addition to its metabolic activity [86,93]. Thus, the AmiC-AmiR pair has been shown to be essential for the effect of natriuretic peptides on anti-biofilm activities, as AmiR overexpression severely impairs the ability of *P. aeruginosa* to form a biofilm [86,93].

This hypothesis of the role of AmiR in *P. aeruginosa* biofilm regulation is supported by the observation that the *amiE* mRNA synthesis is enhanced by 18.3- and 34.4-fold in the biofilm state compared to the planktonic or dispersed state, respectively [94]. In addition, AmiE abundance is increased in 96-hour-old biofilms compared to 48-hour-old biofilms [95] or compared to the planktonic state [96]. It also appears that AmiE is the protein with the highest increased expression among all the proteins identified (8.4-fold) in *P. aeruginosa* biofilm grown for 96 h compared to planktonic cells.

Since biofilm activities are modified by AmiR and the products of the *ami* operon, a direct impact on *P. aeruginosa* pathogenesis was suspected. Indeed, overproduction of AmiE resulted in a total loss of *P. aeruginosa* virulence in a mouse model of acute infection [97]. The involvement of AmiR and members of the *ami* operon in the regulation of various physiological processes strongly suggests that AmiR has more molecular targets than originally identified [85,86,97].

Taken together, these data demonstrate that an RNA-binding antiterminator can have many different functions depending on the physiological state of the bacteria. The data on the *ami* operon in general and AmiR in particular, pave the way for a new perspective on AmiR as a potential regulator of *P. aeruginosa* pathogenicity.

### The NasR antiterminator in Klebsiella oxytoca

In *Klebsiella oxytoca* (sometimes misnamed as *K. pneumoniae*) M5a1, a bacterium belonging to the *Enterobacteriaceae* family, transcriptional antitermination serves as a regulatory mechanism to modulate nitrogen metabolism. This microorganism can use nitrate and nitrite as sole nitrogen sources during aerobic growth. The genes involved in nitrate and nitrite assimilation are organised in an operon, *nasFEDCBA* (Figure 6A), hereafter referred to as the *nasF* operon [98]. This non-catabolic operon shares regulatory similarities with catabolic operons and encodes components essential for nitrate assimilation. Specifically, the *nasF*, *nasE*, and *nasD* genes encode a nitrate and nitrite uptake system [99]. Nitrate is enzymatically converted from nitrite to ammonium through assimilatory nitrate reductases, encoded by *nasC* and *nasA*, and assimilatory nitrite reductase, encoded by *nasB* (Figure 6A). Upstream of *nasF* is *nasR*, which encodes an antiterminator protein [100]. In the presence of nitrate or nitrite, NasR triggers the expression of the *nasF* operon, presumably by facilitating transcription read-through at a terminator within the *nasF* operon leader [101].

The expression of the *nasF* operon in *K. oxytoca* M5a1 is under two levels of regulation. The first control, Ntr control, is a direct result of the cooperative interaction between NtrC, which responds to nitrogen limitation, and RpoN ($\sigma^N$ or $\sigma^{54}$), which directs the binding of RNAP to the conserved -12 (TGC) and -24 (GG) promoter elements (Figure 6A) [101,102]. The sigma factor $\sigma^N$ cannot bind to DNA and initiate transcription on its own. $\sigma^N$-dependent transcription requires the presence of activators that typically bind to sites 80 to 150 bp upstream of the promoter, known as upstream activator sequences (UAS) or enhancer sites. These $\sigma^N$-dependent transcriptional activators are referred to as bacterial enhancer binding proteins (bEBPs) [103,104]. In *K. oxytoca*, NtrC plays a critical role as a bEBP and is activated when *K. oxytoca* senses nitrogen limitation. When activated by phosphorylation, the bEBP NtrC undergoes a conformational change that allows it to bind to a conserved enhancer sequence located upstream of the *nasF* promoter and acts as a transcriptional activator (Figure 6A). It uses a mechanism of ATP hydrolysis to assist RNA polymerase in initiating transcription at promoters controlled by $\sigma^N$ [103]. In addition, the bEBP NtrC has been shown to be involved in the regulation of the transcription of *nasR* which originates from a $\sigma^N$-dependent promoter (Figure 6A). However, the *nasR* upstream region does not contain a UAS for the NtrC [105]. Taken together, this regulatory system ensures that *K. oxytoca* can efficiently switch to nitrate reduction and use this last under conditions where more preferred nitrogen sources (*e.g.*, ammonium) are scarce. The second control of nitrate and nitrite assimilation is governed by the key regulator NasR (Figure 6A). Genetic analyses showed that the expression of the *nasF* operon in response to nitrate/nitrite is specifically mediated by a transcriptional antitermination mechanism, in which NasR interacts with the *nasF* leader region [100,101].

*NasR structure*

In the absence of added nitrate or nitrite, NasR forms dimers in both the crystal and solution states [106]. The NasR protein structure exhibits an organization with two separate domains connected by an elongated linker (Figure 6B). Each subunit within this dimer contains two distinct all-helical domains: an amino-terminal NIT (nitrate- and nitrite-sensing) domain responsible for sensing nitrate/nitrite levels [107] and a carboxyl-terminal ANTAR signalling domain (Figure 6B).

*NIT domain.* The NIT domain found in NasR which spans residues 9 to 287 and encompasses helix α1 to α8 (as shown in Figure 6B), has been detected in various receptor components of signal transduction pathways in a wide range of bacterial species [106,107]. The NIT domain consists of two closely resembling four-helix bundles (Figure 6B), each of which bears structural similarity to periplasmic input domains found in transmembrane receptors such as Tar and NarX. The manner in which these two bundles associate within the NIT domain closely mirrors the homodimeric structures formed by periplasmic input domains. In particular, within the NIT domain, the amino-terminal helix α1′ extends from the packed bundle and interacts with the adjacent monomer.

*Linker region.* The linker region consists of three components: a disordered peptide (residues 288 to 299), a small helical peptide labelled as α9' (residues 302 to 312), and a long helix labelled α9 that is tightly packed against the helical bundle domain (Figure 6B)[106].

*ANTAR domain.* The ANTAR domains play a critical role in modulating various bacterial processes, either through the TCS or through direct interaction with protein sensor domains [13,18]. In *K. oxytoca* M5a1, the ANTAR domain of the NasR protein, spanning residues 333 to 393, shares similarities with ANTAR domains identified in transcriptional antiterminators such as AmiR of *P. aeruginosa* and Rv1626 of *Mycobacterium tuberculosis*. The ANTAR domain (Figure 6B) comprises a triangular configuration of a three-helix bundle (referred to as helix α10 to α12), similar to the arrangement observed in the ANTAR domains of AmiR and Rv1626 [83,108]. Notably, the first helix (α10) of the ANTAR domain is significantly longer than the other two and is oriented perpendicular to the linker-derived helix α9 (Figure 6B) [106]. The ANTAR signalling domain of NasR targets a regulatory element known as the "antiterminator" intrinsic secondary structure located within the leader region of the *nasF* operon mRNA transcript [106,109].

*NasR antitermination mechanism*

In *K. oxytoca* M5a1, the transcriptional antitermination regulatory mechanism of NasR has been extensively studied [101,109,110]. The NasR-mediated antitermination occurs after its interaction with the *nasF* leader RNA. The NasR target is a two-hairpin RNA motif in the *nasF* leader region. Specifically, each hairpin structure is capped by a terminal hexanucleotide stem-loop containing A and G residues at the first and fourth positions, which are essential for NasR binding. A recent study has proposed the molecular mechanism by which NasR couples its ligand signal to its RNA binding activity [110]. In the absence of the nitrate signal, the NasR antiterminator forms a constitutive dimer that is unable to bind RNA by adopting an autoinhibited conformation (Figure 6C, left). However, in the presence of nitrate, the autoinhibitory conformation is disrupted: NasR adopts an alternative configuration by releasing the key RNA recognition residues of the ANTAR domain to bind the hairpin RNA structures that contribute to the antitermination (Figure 6C, right). Binding of nitrate to the NIT domain could induce structural transitions in the ANTAR domain to adopt an alternative dimer configuration, allowing the release of its RNA binding surface [110].

*NasR antitermination and pathogenicity*

Nitrogen is a critical nutrient for bacterial growth and is readily available to bacteria in various environments in the form of ammonium, nitrate, and nitrite. When ammonium is scarce, many bacterial strains, including *K. oxytoca* M5a1, which possess an assimilatory nitrate/nitrite reductase pathway, can use them as the sole nitrogen source for growth and to maximise their fitness in the host environment [111]. During infection, *Klebsiella* species can use a variety of virulence factors for survival and immune evasion [112]. In particular, cytotoxins, capsular polysaccharide (K antigen), LPS (containing O antigen), fimbriae, outer membrane proteins, and determinants for iron acquisition and nitrogen source utilisation are common virulence factors in *K. pneumoniae* and *K. oxytoca* [112,113]. In addition, some studies reported that clinical isolates of *K. oxytoca* collected from patients with antibiotic-associated haemorrhagic colitis (AAHC) [114] and colorectal cancer [115] were able to form moderate biofilms that prevented efficient eradication of infection.

The relationship between nitrogen metabolism regulation and bacterial pathogenicity is complex. Bacterial pathogens often encounter varying nitrogen environments during infection, and their ability to adapt and use nitrogen sources can significantly affect their virulence and survival within the host. Changes in nitrogen availability within the host can impact the expression of host defence mechanisms and the ability of bacteria to evade immune responses. Adequate nitrogen supports the synthesis of amino acids, which are key players in the modulation of immune responses. These basic building blocks promote effective immune cell function, proliferation, and differentiation through the production of antibodies, cytokines, nitrogen-containing molecules (such as nitric oxide), and the activation of T and B lymphocytes, natural killer cells, and macrophages [112,116–119]. Conversely, nitrogen deficiency or imbalance can compromise immune responses, increasing susceptibility to infection. Thus, maintaining nitrogen balance is critical for immune homeostasis. Furthermore, it has been demonstrated that nitrogen metabolism can also influence biofilm formation, a critical aspect of bacterial pathogenesis. Overall, nitrogen metabolism plays a central role in shaping bacterial pathogenicity by influencing nutrient acquisition, virulence factor production, host-pathogen interactions, and biofilm formation [111]. The relationship between bacterial pathogenicity and nitrogen metabolism is intricate. It is difficult to delineate the boundaries between the two, however, it is apparent that they are interrelated. Consequently, we can speculate that the transcriptional antitermination regulatory mechanism of NasR involved in nitrogen metabolism may directly or indirectly affect the pathogenicity of *K. oxytoca* M5a1, thereby enabling its persistence in the host environment, evasion of immune responses, and resistance to antimicrobial treatments. Therefore, a comprehensive understanding of the molecular regulatory mechanisms that govern this complex relationship is essential for the development of effective strategies to combat infectious diseases.

### *RNA-binding antiterminators in* **Mycobacterium tuberculosis** *and* **Enterococcus faecalis**

ANTAR antiterminators bind to RNAs with a conserved dual stem-loop structure found in the 5'UTRs of specific mRNAs. As described for AmiR and NasR in Gram-negative bacteria, this binding stabilises a transcription antiterminator, thus allowing transcription. In some species of the phyla *Bacillota* and *Pseudomonadota*, histidine kinases have been observed to phosphorylate the ANTAR response regulator protein, and then activate ANTAR for RNA binding.

An extensively studied ANTAR antiterminator involved in the switch from commensal life to pathogenesis is the EutV protein in *E. faecalis*. This bacterium can use ethanolamine as a nutrient source, providing a selective advantage in the intestinal tract and contributing to its pathogenic potential. The complex regulatory model of the *eut* ethanolamine utilisation operon involves a cobalamin-dependent riboswitch and the TCS EutW-EutV [120]. Ethanolamine acts as a signalling ligand that induces autophosphorylation of the histidine kinase EutW. This enzyme in turn phosphorylates the response regulator EutV, following the consensual model of TCS, except that the C-terminal domain of EutV contains an ANTAR domain like the AmiR protein in *P. aeruginosa*. Thus, the response regulator binds to the *eut* 5'UTR, on an RNA sequence upstream of a Rho-independent terminator. Antitermination of transcription occurs by preventing the formation of the terminator hairpin. Termination at a coenzyme B12-dependent riboswitch in the presence of cobalamin also increases the efficiency of both transcription initiation and antitermination by EutV [120,121]. A model of the protein/RNA interaction has been developed following a successful structural characterisation [14]. This regulation of ethanolamine catabolic enzymes has also been described in pathogens such as *Listeria* and *Clostridium* [16].

A large number of ANTAR proteins have also been identified in the mycobacterial genome, with a significant overrepresentation compared to *Pseudomonas* or *Clostridium* [122], and a similar structure to ANTAR antiterminator proteins [123]. ANTAR RNA target sequences often overlap with the ribosome binding site or the translation start site of the regulated genes, resulting in translation inhibition. In mycobacteria, the Rv1626 ANTAR protein is constitutively active and regulation occurs at its activating histidine kinase Rv3220c. Targets of ANTAR antiterminator have been shown to be genes involved in lipid and redox pathways, with important effects on virulence in *M. tuberculosis*, for example [122]. The link between ANTAR proteins and lipid catabolism has also been found in *M. smegmatis*, *M. vaccae* and *M. vanbalenii*, but also in *E. coli* and *Rhodococcus* spp. Both pathogenic and saprophytic mycobacteria use host/environment lipids for growth and survival, and resistance to oxygen and nitrogen species provides advantages for survival in the host. This underlines the important role of ANTAR antiterminator regulation in actinobacteria.

## Other RNA-binding antiterminators

### *The HutP antiterminator in* Bacillus subtilis

L-histidine is an amino acid that can be used by *B. subtilis* as a source of energy, carbon, and nitrogen for growth, although its degradation is tightly regulated due to its costly synthesis in the cell [124]. The *B. subtilis hutPHUIGM* operon (histidine utilisation), hereafter referred to as the *hutP* operon, encodes enzymes for the transport and catabolism of L-histidine, as well as an RNA-binding protein that positively regulates its expression [125–127]. The *hutM* gene encodes a histidine transporter, while the *hutHUIG* genes are involved in histidine utilisation. The pathway of histidine catabolism is highly conserved among bacteria, although two different routes can be distinguished. In *B. subtilis*, histidine is deaminated by HutH histidinase to give urocanate, which is further hydrated by HutU urocanase to give imidazolone propionate; the latter is ring-cleaved by HutI hydrolase to give formiminoglutamate, which is finally hydrolysed to formamide and glutamate by HutG [124,125,128]. The regulatory elements consist of *hutP*, encoding a protein involved in the antitermination of the entire *hutPHUIGM* transcript, and a transcription terminator, located between *hutP* and *hutH*, to whose mRNA hairpin HutP binds [129–131].

*HutP structure*

The HutP from *B. subtilis* forms a homohexameric complex resembling a flattened cylinder, with three individual dimers linked by three-fold symmetry [132]. According to the solved crystal structure of HutP, each monomer consists of four antiparallel β-strands forming a β-sheet, with two α-helix on the front side and two more at the back, and five hairpin regions (L) (Figure 7A). A crystal structure of the HutP hexamer revealed that it recognises and binds a single-stranded RNA molecule on each side of the cylinder in a triangular conformation (Figure 7B). However, prior binding of both L-histidine and $Mg^{2+}$ ions to each HutP monomer is required to activate the RNA-binding activity of the hexamer [133]. The L-histidine and $Mg^{2+}$ ligands are associated with each other and buried inside the HutP cylinder at the interface between two HutP monomers, while the RNA molecules are bound to both outer surfaces. Therefore, the structure of the HutP quaternary complex contains a protein homohexamer with six L-histidines, six $Mg^{2+}$ ions, and two bound single-stranded RNA molecules [132,134] (Figure 7B).

*HutP antitermination mechanism*

By contrast to *B. subtilis*, the *hut* genes of Gram-negative microorganisms (*i.e.* enteric bacteria and several species of *Pseudomonas*) do not reside in a single operon [124]. In addition, a *hutT* gene encoding a histidine/urocanate transporter instead of the aforementioned histidine transporter (HutM), and a Hut-specific repressor HutC instead of the HutP antiterminator have been identified in Gram-negative bacteria. Therefore, regulation of the *hutP* operon in enteric bacteria and pseudomonads is quite different from that observed in *B. subtilis*, with HutC-mediated repression in the former and HutP-mediated antitermination in the latter [124,129–131].

Hut enzymes in *B. subtilis* are induced by L-histidine, rather than urocanate as in enteric bacteria and pseudomonads [124,135], and are repressed by rapidly metabolised carbon sources such as glucose and the presence of amino acids [135,136]. L-histidine-mediated induction of the *hutP* operon in *B. subtilis* is regulated by an antitermination mechanism [124]. CCR is mediated by the CcpA protein, which binds to the *cre* site located in the middle of the *hutP* gene [124,137], whereas transcriptional repression in response to amino acid availability is controlled by the CodY protein, which acts at the *hutO_A* site located immediately downstream of the *hutP* operon promoter (Figure 7C) [130,138].

Expression regulation of the histidine utilisation system in *B. subtilis* by antitermination mechanisms has been reviewed previously [124,127,139–141]. In the presence of the inducer, six molecules of L-histidine bind to the inactive HutP hexamer, which undergoes a conformational change. Then, six $Mg^{2+}$ ions also bind to the HutP-L-histidine structure, triggering the active conformation of the HutP-L-histidine-$Mg^{2+}$ complex for RNA recognition [132,142–144]. When activated by the coordination of L-histidine and $Mg^{2+}$ ions, HutP binds specifically to the target sequence within the terminator RNA located within the *hutP-hutH* intercistronic region (Figure 7C) and initiates the destabilisation process that compromises the terminator hairpin structure.

Specifically, the hexameric activated HutP interacts with two clusters (designated site I and II) within the target RNA, each one consisting of three nucleobase-adenine-guanine (NAG) motifs separated by two to four unconstrained spacer nucleotides, with a 20 pb GC-rich linker region between binding sites I and II (Figure 7D) [127,145–147]. Each HutP monomer recognises one NAG triplet of the terminator RNA, so that the six NAG regions are bound by the six protein subunits, three motifs (one cluster) on each surface of the hexamer, with the 20-bp linker separating each cluster [132,134]. Thus, under conditions of excess L-histidine, the activated HutP hexamer first accesses to the NAG-rich site I within the nascent *hut* transcript by using one of its two RNA-binding surfaces, and then binds to site II using the opposite RNA-binding side to mediate the conformational changes of the terminator RNA hairpin, thus preventing premature transcription termination and allowing read-through of the terminator [147].

In addition to the antitermination regulation of the Hut enzymes, HutP can also activate the promoter of the *B. subtilis hutP* operon. However, the mechanism by which HutP activates transcription from the *hutP* promoter remains unknown [124].

*HutP homologs in pathogenic bacteria*

Although *B. subtilis* cannot be considered as human pathogen, orthologs of the HutP regulatory protein have been detected in closely related species responsible for disease, such as *Bacillus cereus* [148] and *Bacillus anthracis* [149]. It has been proposed that the Hut system is required to generate energy using histidine as sole carbon and energy sources and to provide a critical nitrogen source during infection[150]. Therefore, it can be hypothesised that HutP-mediated antitermination may promote pathogenesis by allowing histidine bioavailability.

**Conclusion**

Antitermination regulation is an important strategy used by bacteria to adapt to their environment. In particular, pathogens need to optimize their nutrient acquisition and metabolism, involving RNA-binding antiterminators described in this review, with an implication for virulence traits. Moreover, the interaction dynamics between antiterminators and RNA represent excellent targets for the development of alternative and innovative drugs. These targets can be selectively modulated through direct and allosteric inhibition of enzymatic activity or inhibition of protein-protein or protein-nucleic acid interactions. Therefore, a comprehensive understanding of the molecular regulatory mechanisms that govern the complex relationship between transcriptional antitermination and pathogenicity is essential for the development of effective strategies to combat infectious diseases and opens up interesting perspectives.

**Conflict of Interest**

The authors declare that the study was conducted in the absence of any commercial or financial relationship that could be construed as a potential conflict of interest.

No funding was received

The data that support the findings of this study are openly available in arXiv repository at http://doi.org/[doi], reference number [reference number].

1    **Table 1:** List of RNA-binding antiterminators found in model bacteria

| RNA-binding antiterminators | Model bacteria | RNA target | Target operon | Main Targets | Involvement in pathogenic cellular process | References |
|---|---|---|---|---|---|---|
| *Antiterminators with RNA-binding CAT domain* | | | | | | |
| SacY | *Bacillus subtilis* | RAT element | *sacB, sacXY* | Sugar metabolism (sucrose utilisation) | *non pathogenic bacteria* | 52 |
| SacT | *Bacillus subtilis* | RAT element | *sacPA* | Sugar metabolism (sucrose utilisation) | *non pathogenic bacteria* | 48 |
| LicT | *Bacillus subtilis* | RAT element | *licTS, bglPH* | Sugar metabolism (β-glucans utilisation) | *non pathogenic bacteria* | 66 |
| LicT | *Streptococcus pyogenes* *Streptococcus gordonii* | RAT element | *bglPB* | β-glucosides utilisation | Regulation of virulence-related genes involved in blood dissemination, biofilm formation, streptolysin S-mediated haemolysis and localized ulcer lesion formation. *in vitro* adhesion, biofilm formation, growth, and *in vivo* colonization | 76, 77 |
| BglG | *Escherichia coli* | RAT element | *bglGFB* | Sugar metabolism (β-glucosides utilisation) | Induced *in vivo* during sepsis | 19, 45 |
| NagY | *Enterococcus faecalis* | RAT element | *nagYE, hylA* | Degradation of glycosaminoglycans | Modulation the expression of virulence factor | 12 |
| *Antiterminators with RNA-binding ANTAR domain* | | | | | | |
| AmiR | *Pseudomonas aeruginosa* | Dual hairpins | *ami* | Amide metabolism | Impairment biofilm formation; loss of virulence of in a mouse model of acute infection | 85, 86, 92, 93 |
| NasR | *Klebsiella oxytoca* | Dual hairpins | *nasF* | Nitrate assimilation | Direct or indirect involvement in : nutrient acquisition; virulence factor production; host-pathogen interactions; biofilm formation; Activation of T and B lymphocytes, natural killer cells and macrophages | 99, 100, 101, 102 |
| EutV | *Enterococcus faecalis* | Dual hairpins | *eut* | Ethanolamine utilisation | Host survival and virulence | 120, 121 |
| Rv1626 | *Mycobacterium tuberculosis* | Dual hairpins | | Unknown | Modulation of lipid and redox pathways important for virulence, growth and survival | 123 |
| **Other antiterminator** | | | | | | |
| HutP | *Bacillus subtilis* *Bacillus cereus* *Bacillus anthracis* | Dual hairpins | *hutPHUIGM* | Amino acid metabolism (histidine utilisation) | Histidine scavenging in the host | 147 |

**Figures**

A

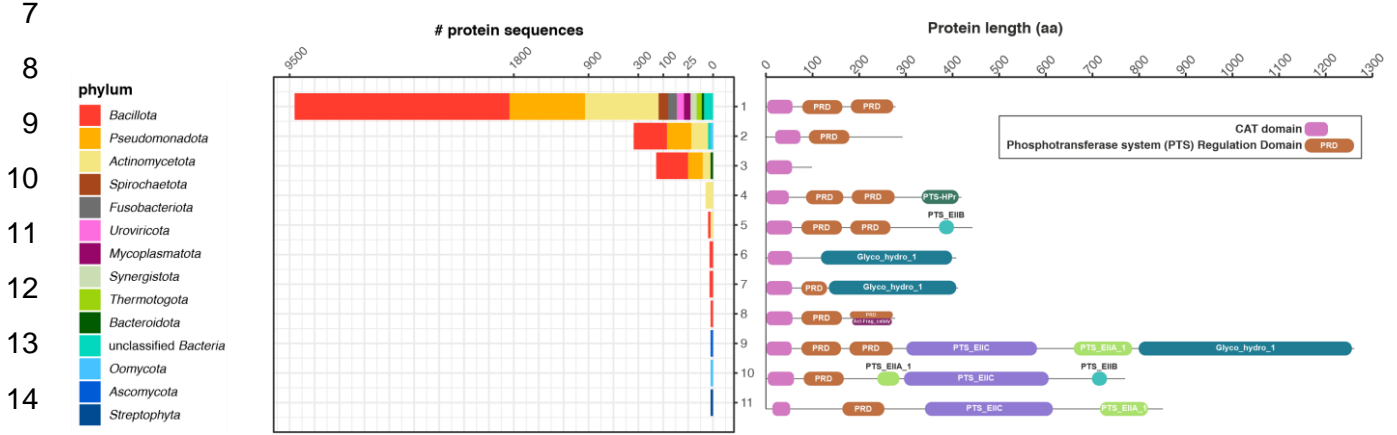

B

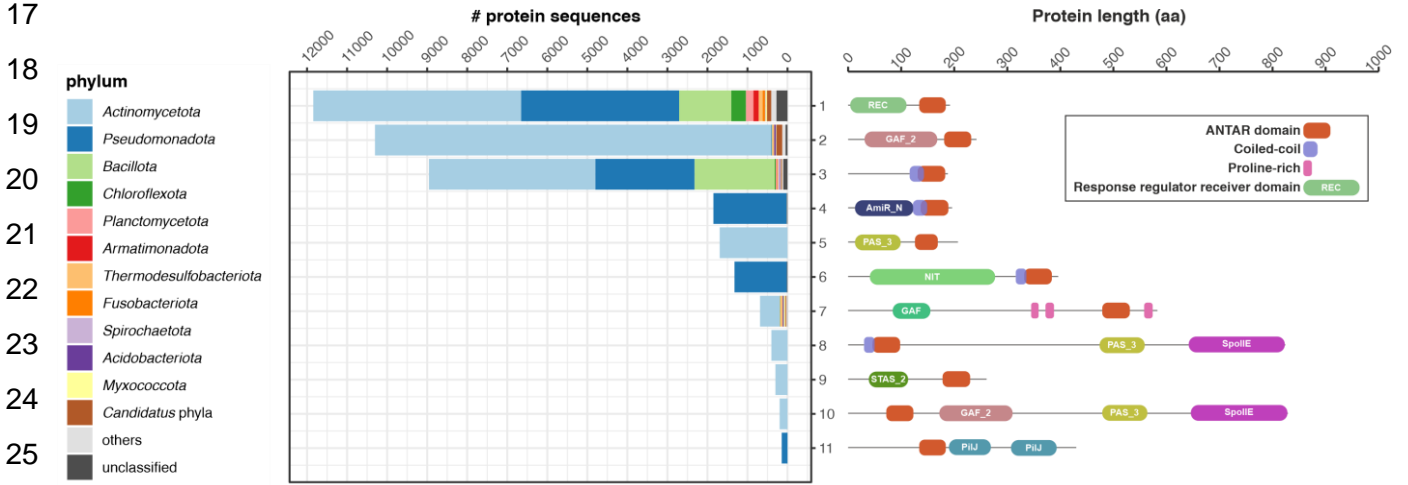

**Figure 1.** Distribution of most abundant RNA-binding CAT and ANTAR domain architectures among bacterial phyla. Different domains are indicated in different colors. CAT and ANTAR proteins and their domain architectures were retrieved from Pfam v37.0 (ID: PF03123 and PF03861, respectively). Out of the 21 and 149 domain architectures found in CAT and ANTAR proteins, respectively, the eleven most represented ones were selected and their taxonomy diversity was analyzed according to NCBI Taxonomy Database. Plots were generated in R using ggplot2 v3.5.1 library. Abbreviations: PRD, Phosphotransferase system Regulation Domain; PTS-HPr, histidine-containing phosphocarrier protein of the phosphotransferase system; PTS_EIIB, Enzyme IIB domain of the phosphotransferase system; Glyco_hydro_1, Glycosyl hydrolase family 1; Act-Frag_cataly, Actin-fragmin kinase catalytic domain; PTS_EIIC, Enzyme IIC domain of the phosphotransferase system; PTS_EIIA_1, Enzyme IIA

1 domain of the phosphotransferase system; REC, Response regulator receiver domain; GAF_2, cGMP-specific phosphodiesterases, adenylyl cyclases and FhlA domain 2; AmiR_N, N-terminal domain of aliphatic amidase regulator; PAS_3, Per-Arnt-Sim fold 3; NIT, Nitrate and nitrite sensing; GAF, cGMP-specific phosphodiesterases, adenylyl cyclases and FhlA domain; SpoIIE, Stage II sporulation protein E; STAS_2, Sulphate Transporter and AntiSigma factor antagonist domain; PilJ, Type IV pili methyl-accepting chemotaxis transducer N-terminal domain.

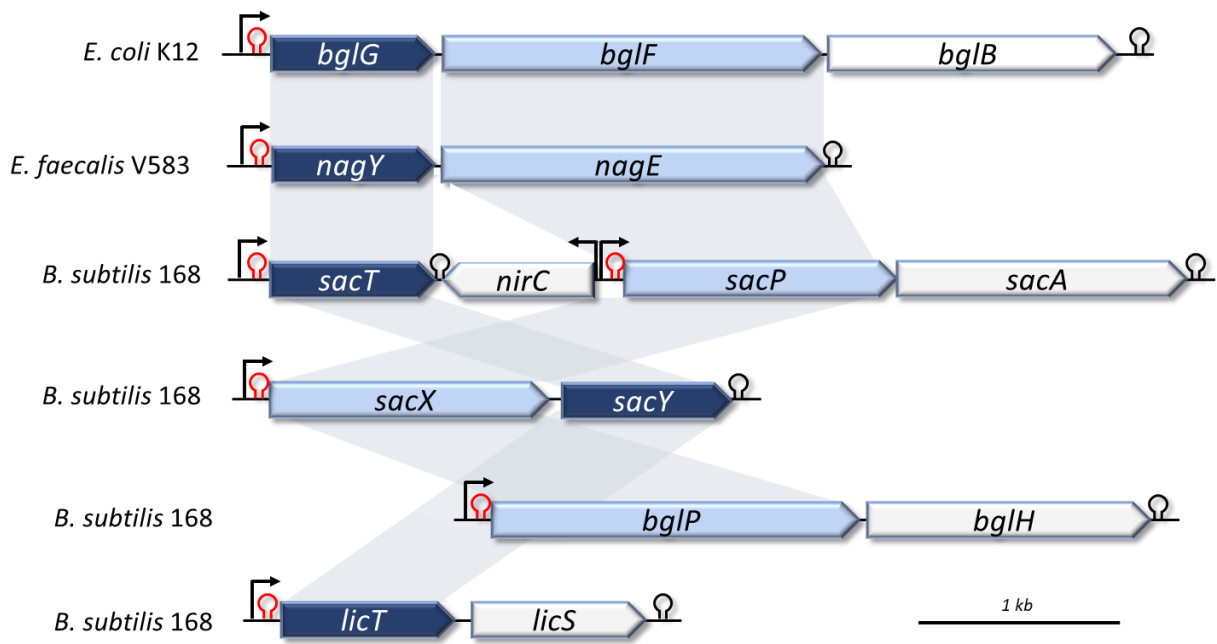

**Figure 2.** Operonic structure encoding BglG in *E. coli*, NagY antiterminator in *E. faecalis*, and SacT/SacY/LicT antiterminators and the LicT-regulated *bglPH* operon in *B. subtilis*. Genes encoding antiterminators and carbohydrate transporters are represented in dark and light blue, respectively. Antiterminators binding sequences are displayed in red.

55

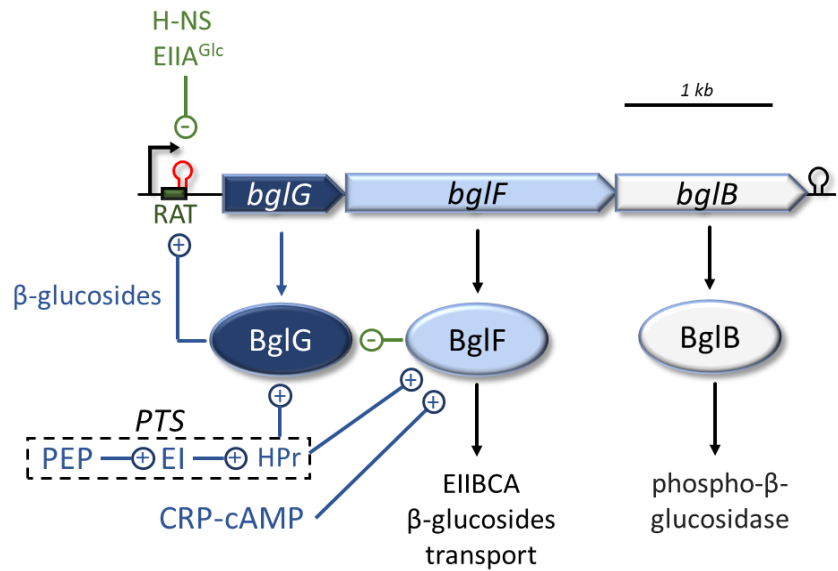

**Figure 3.** BglG antitermination mechanism in *E. coli*. BglG activity is dependent on its phosphorylation steps, involving HPr from the phosphotransferase system (PTS), and the CRP-cAMP complex. The first step of the phosphorylation cascade catalysed by the PTS is the autophosphorylation of Enzyme I (EI) with phosphoenolpyruvate (PEP). The phosphate is then transferred to HPr which phosphorylates BglF (EIIBCA). In the absence of β-glucoside, BglG is inactivated by BglF. The *bgl* operon transcription is initiated by RNAP but stops at the red RNA hairpin. In the presence of β-glucoside, BglF phosphorylates the carbohydrate during its uptake, and BglG is activated. After dimerization, BglG binds the Ribonucleic Antiterminator sequence (RAT), promoting antitermination, and expression of the *bgl* operon. Transcription termination occurs at the end of *bglB* at the terminator represented by a black stem-loop. H-NS and EIIA$^{Glc}$ also provides regulation on the *bgl* operon.

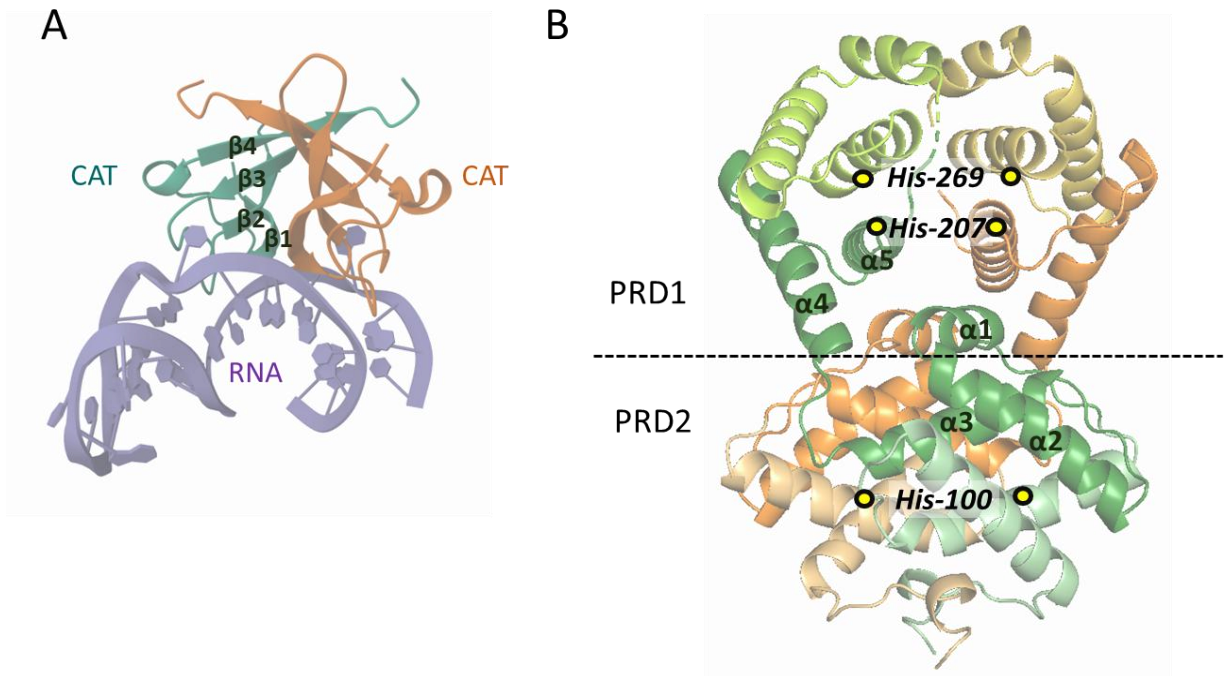

**Figure 4.** Structure of the SacT/SacY/LicT CAT and PRD domains of *B. subtilis* antiterminators. (A) Dimer of the CAT (RNA-binding domain) in complex with its hairpin RNA target (PDB 1L1C). The two monomers are represented in orange and green, and the double-strand RNA is displayed in purple, β-sheets number of one of the 2 CAT domains represented are indicated. (B) Dimers of PRD domains with the phosphorylatable histidines (position in LicT) (PDB 1H99). The two monomers are represented in orange and green. α-helix number are indicated for one of the two monomers.

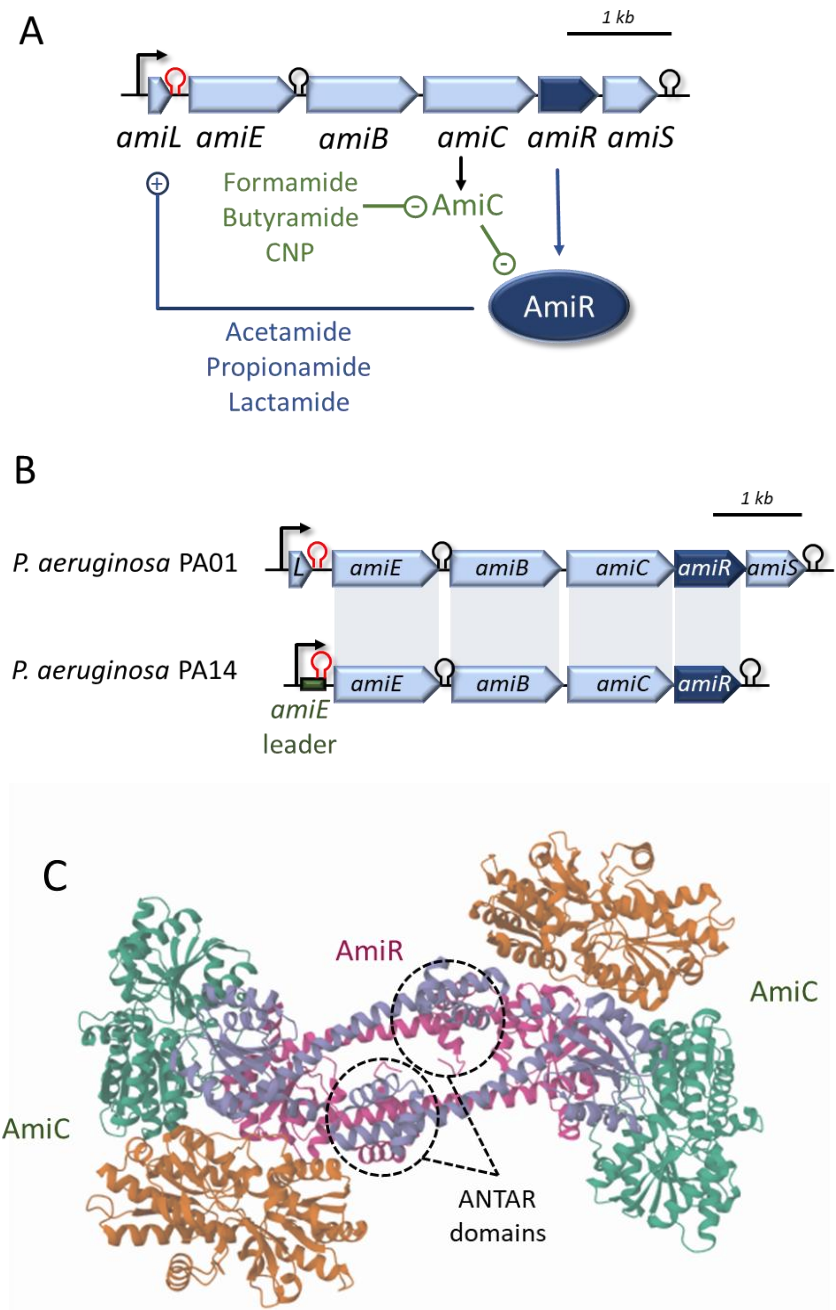

**Figure 5.** AmiR antitermination mechanism in *P. aeruginosa*. (A) In the PAO1 strain, AmiC represses AmiR activity, but in presence of acetamides AmiR binds on *amiL* and antitermination occurs at the red RNA hairpin. (B) Operon structure in *P. aeruginosa* PAO1 and PA14 strains, adapted from *www.pseudomonas.com*. In PA14 strain, *amiL* is designated as *amiE* leader. Transcription terminators are indicated by a black stem-loop. (C) Amidase sensor AmiC of the amidase operon of *P. aeruginosa* complexed with the AmiR antiterminator (PDB 1QO0). AmiR monomers are displayed in pink and purple, and AmiC proteins in green and orange. ANTAR (RNA-binding domain) of AmiR proteins are indicated.

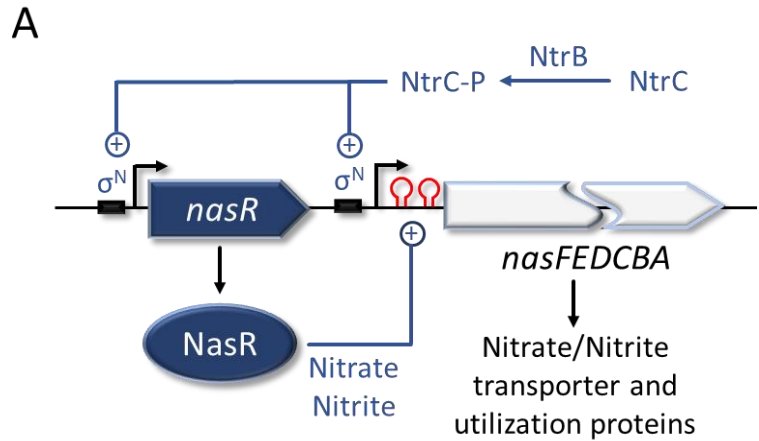

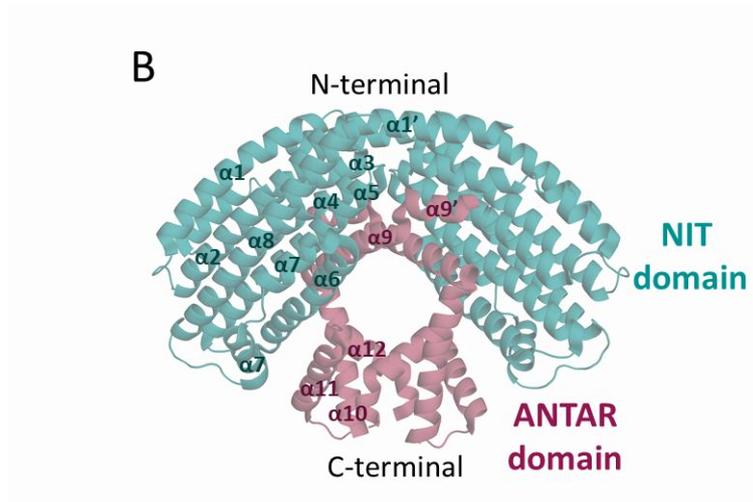

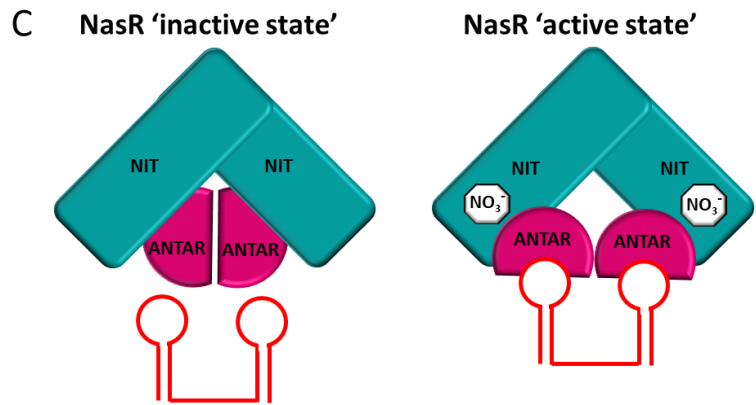

**Figure 6.** NasR antitermination mechanism in *K. oxytoca* M5a1. (A) The regulation of the *nasF* operon (*nasFEDCBA*) required for nitrate and nitrite assimilation is tightly controlled and involves a combination of environmental signals (such as nitrate/nitrite availability) and genetic regulatory circuits. The primary mechanism of regulation involves the Ntr system and $\sigma^N$ ($\sigma^{54}$) and controls the expression of the *nasF* operon and *nasR* transcription. The second mechanism involves the NasR antiterminator which is typically involved in sensing nitrate levels. When nitrate is present, NasR activates the transcription of the *nasF* operon by binding to the 2 red RNA hairpins in the promoter region of the operon (see C). (B) Structure of the NasR transcriptional antiterminator (PDB 4AKK). NasR is a dimeric protein with two domains: the NIT (nitrate- and nitrite-sensing) domain (green) and the ANTAR (AmiR and NasR Transcription Antitermination Regulator) domain (purple). The NIT domain detects nitrate/nitrite levels, while the ANTAR domain regulates transcription of the *nasF* operon by interacting with its mRNA leader region. α-helix are indicated. (C) NasR mediates transcription antitermination by binding to a two-hairpin RNA motif in the *nasF* leader region. In the absence of nitrate, NasR forms an autoinhibited dimer (NasR 'inactive state'), but in the presence of nitrate, it undergoes a conformational change, enabling RNA binding and antitermination (NasR 'active state').

A 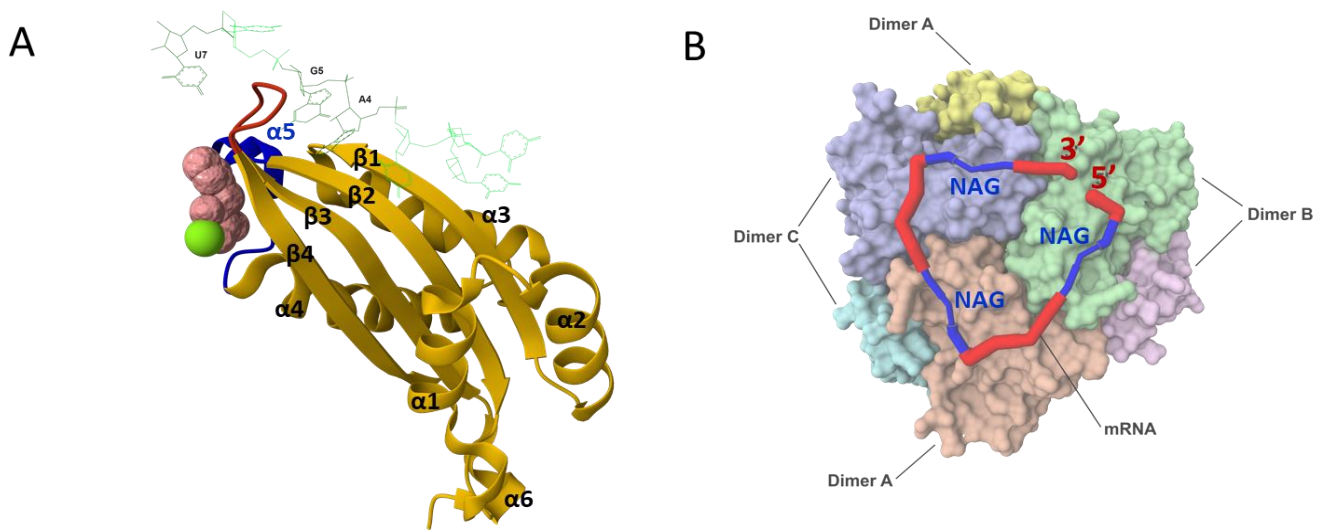 B

109

C 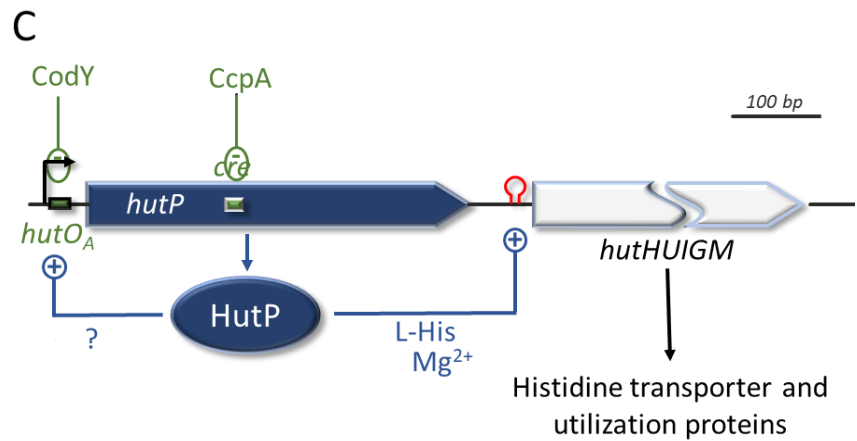

110

D 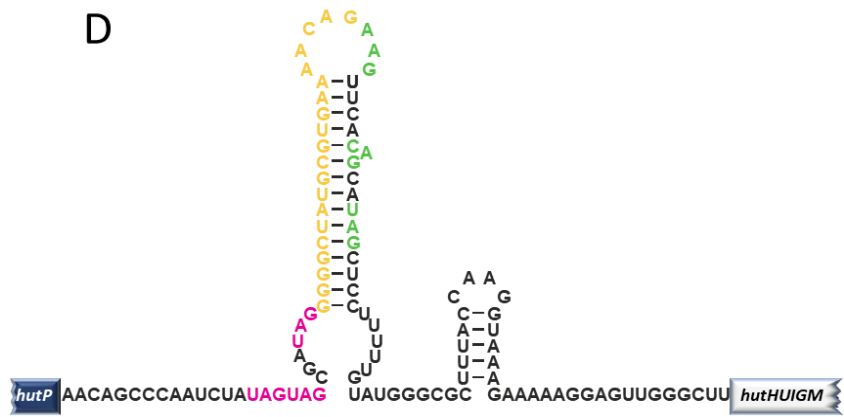

111

**Figure 7.** HutP antitermination mechanism in *B. subtilis* (A) Detail of one HutP protein monomer (PDB 3BOY) interacting with the RNA, histidine, and Mg²⁺. The HutP protein is displayed in yellow, with six α-helix and four antiparallel β-strands arranged in the order α₁-α₂-

β1-α3-α4-α5-β2-α6-β3-β4. The protein helix (α5) and stem-loops suggested to be involved in the structural communication between the ligand-binding pocket and the RNA are marked in blue and red, respectively. The RNA is shown in green, with protein-bound nucleotides highlighted in dark green and labelled. The surface of the histidine ligand is shown in pink and the $Mg^{2+}$ ion as a green sphere. (B) Top view of the structure of the HutP quaternary complex (PDB 3BOY). Each monomeric protein unit forming a homohexamer (with three individual dimers) is represented with a different colour, the mRNA segment is represented as a red tube, and the blue residues highlight the three specific NAG binding sites (Nucleobase-Adenine-Guanine). The ligand-binding pockets and the second bound RNA are not visible in this view. (C) Schematic representation of the *hutPHUIGM* operon structure and regulation. Binding sites for regulatory proteins (CodY on HutO$_A$ and CcpA on *cre*-box) are indicated by green boxes. In presence of $Mg^{2+}$ and L-Histidine, HutP is activated and binds to the antiterminator binding sequence represented by a red RNA hairpin, to allow the *hutHUIGM* operon expression (D) HutP-target terminator RNA structure. NAG recognition motifs of sites I and II are highlighted in pink and green, respectively, and the 20-bp linker region between the two HutP binding sites is shown in orange.